\documentclass{PoS}

\title{Theoretical Results on Neutrinos}

\ShortTitle{Theoretical Results on Neutrinos}

\author{\speaker{Shun Zhou}%
         \thanks{This work was supported in part by the Innovation Program of the Institute of High Energy Physics under Grant No. Y4515570U1, by the National Youth Thousand Talents Program, and by the CAS Center for Excellence in Particle Physics (CCEPP).}\\
        Institute of High Energy Physics, Chinese Academy of Sciences, Beijing 100049, China\\
        Center for High Energy Physics, Peking University, Beijing 100871, China \\
        E-mail: \email{zhoush@ihep.ac.cn}}

\abstract{In this talk, I first summarize our current knowledge about the fundamental properties of neutrinos and emphasize the remaining unsolved problems in neutrino physics. Then, recent theoretical results on neutrino mass models are introduced. Different approaches to understanding tiny neutrino masses, lepton flavor mixing and CP violation are presented. Finally, I report briefly some new progress in the studies of astrophysical neutrinos, including keV sterile neutrinos, supernova neutrinos and ultrahigh-energy cosmic neutrinos.}

\FullConference{ International Symposium on Lepton Photon Interactions at High Energies\\
		 17-22 August 2015\\
		 University of Ljubljana, Slovenia}

\begin{document}

\section{Introduction}

\begin{figure}[t]
\centering
\includegraphics[height=2.8in]{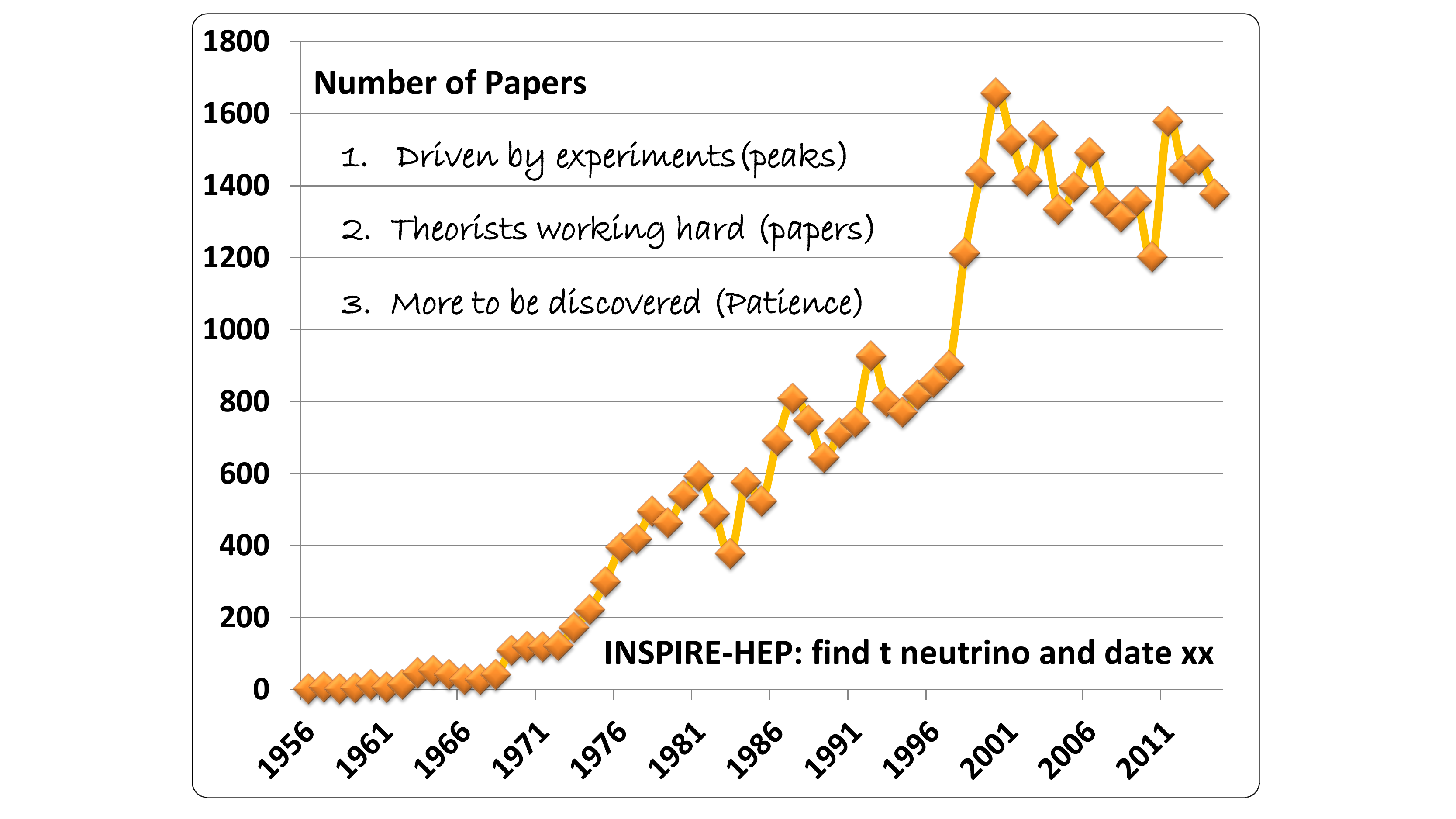}
\caption{\it The number of neutrino papers has been shown as a function of the year of publication, where the data have been obtained from INSPIRE-HEP by searching for the keyword "neutrino" in titles.}
\label{fig:numbers}
\end{figure}

Although the neutrino was first conjectured in 1930 by the great theorist Wolfgang Pauli, thus far every important step forward in neutrino physics has been led by experimentalists. As many other authors did before, I search in the INSPIRE-HEP database for the papers with "neutrino" in their titles, and plot a curve of the number of neutrino papers with respect to the year of publication. It is straightforward to observe from Fig.~\ref{fig:numbers} that there are many peaks along the gradually increasing curve, which are stimulated by groundbreaking discoveries in neutrino physics. 

In 1956, Clyde Cowan and Fredrick Reines detected $\overline{\nu}^{}_e$ from nuclear reactors for the first time~\cite{Cowan:1992xc}, signifying the birth of neutrino physics. In 1962, Leon Lederman, Melvin Schwartz and Jack Steinberger proved the existence of a different flavor of neutrino, namely $\nu^{}_\mu$ and $\overline{\nu}^{}_\mu$, in nature~\cite{Danby:1962nd}. In 1968, Raymond Davis, Jr. discovered solar neutrinos $\nu^{}_e$ from nuclear fusions in the Sun, and found the discrepancy between the experimental observation~\cite{Davis:1968cp} and the theoretical prediction from the standard solar model~\cite{Bahcall:1968hc}. In 1987, the neutrino burst from SN 1987A, a core-collapse supernova  in the Large Magellanic Clound, was observed in Kamiokande-II, IMB and Baksan experiments~\cite{Hirata:1987hu}. In 1998, the Super-Kamiokande experiment confirmed the disappearance of upward-going $\nu^{}_\mu$ and $\overline{\nu}^{}_\mu$, and provided strong evidence for atmospheric neutrino oscillations~\cite{Fukuda:1998mi}. In 2002, the Sudbury Neutrino Observatory (SNO) experiment measured both $\nu^{}_e$ and non-electron neutrino fluxes, and demonstrated neutrino flavor conversions as a solution to the long-standing solar neutrino problem~\cite{Ahmad:2002jz}. In 2012, the Daya Bay experiment observed the disappearance of reactor neutrinos at a short baseline and found that the smallest leptonic flavor mixing angle $\theta^{}_{13}$ is relatively large~\cite{An:2012eh}, which was later confirmed by both RENO~\cite{Ahn:2012nd} and Double Chooz~\cite{Abe:2012tg} experiments.

The most important progress in the last two decades should be the discovery of neutrino oscillations,\footnote{The 2015 Nobel Prize in Physics was awarded jointly to Takaaki Kajita and Athur B. McDonald for the discovery of neutrino oscillations in the Super-Kamiokande and SNO experiments. In addition, the 2016 Breakthrough Prize in Fundamental Physics has been shared by the major neutrino oscillation experiments and their leaders: Yifang Wang and Kam-Biu Luk for Daya Bay, Atsuto Suzuki for KamLAND, Koichiro Nishikawa for K2K and T2K, Arthur B. McDonald for SNO, Takaaki Kajita and Yoichiro Suzuki for Super-Kamiokande. } which indicate that neutrinos are massive particles and lepton flavors are significantly mixed. The experimental results are crucially important for the whole area to advance considerably, but the research activities are actually dominated by theoretical studies on neutrinos, showing up as peaks in Fig.~\ref{fig:numbers}.  Therefore, it is also worthwhile to mention a few seminal works, some of which were inspried by experimental results and some not.  

In 1930, Pauli proposed the neutrino in order to rescue the law of energy-momentum conservation and explain the continuous energy spectrum of electrons from beta decays. In 1933, based on Pauli's neutrino hypothesis, Enrico Fermi put forward his famous effective theory of beta decays,  and suggested a possible determination of neutrino masses from the shape of energy spectrum near the endpoint~\cite{Fermi:1933jpa}. In 1937, Ettore Majorana considered the possibility that particles could be their own antiparticles, which are now called Majorana particles~\cite{Majorana:1937vz}. Later, in 1939, Wendell Furry noticed a remarkable difference between Majorana and Dirac neutrinos that neutrinoless double-beta decays $N(Z, A) \to N(Z+2, A) + 2 e^-$ could happen only in the former case~\cite{Furry:1939qr}. In 1957, after the discovery of $\overline{\nu}^{}_e$ from nuclear reactors, Bruno Pontecorvo made an interesting analogue between the mixings of neutral $K^0$-$\overline{K}^0$ and $\nu^{}_e$-$\overline{\nu}^{}_e$ systems, and postulated a possible transition $\nu^{}_e \leftrightarrow \overline{\nu}^{}_e$~\cite{Pontecorvo:1957cp}. Even without knowing the discovery of $\nu^{}_\mu$ in 1962, Ziro Maki, Masami Nakagawa and Shoichi Sakata predicted the flavor conversion $\nu^{}_e \leftrightarrow \nu^{}_\mu$ in a theoretical model of elementary particles~\cite{Maki:1962mu}. Nowadays, the lepton flavor mixing matrix is named after those four theorists as PMNS matrix. In 1978, Lincoln Wolfenstein realized that the coherent forward scattering of neutrinos with matter could significantly modify neutrino oscillation phenomena~\cite{Wolfenstein:1977ue}. In 1985, Stanislav Mikheyev and Alexei Smirnov discovered that such a modification could enhance resonantly neutrino flavor mixing and lead to a substantial flavor conversion~\cite{Mikheev:1986gs}. The Mikheyev-Smirnov-Wolfenstein (MSW) matter effects help us solve the solar neutrino problem via neutrino oscillations.

Most of the above theoretical ideas were indispensable for us to explain the experimental observations and understand the fundamental properties of neutrinos. In this talk, I shall summarize many recent ideas about the origin of neutrino masses and flavor mixing. While great experimental efforts are now being devoted to answering a number of basic questions in neutrino physics, I hope some of those ideas will prove to be important and more new ideas are to come.

\section{Fundamental Properties of Massive Neutrinos}
 In the Standard Model of elementary particles (SM), there are three generations of massless neutrinos, participating in the weak interactions, and they are of spin one-half and have no electric charges. Now we know, the only difference is that neutrinos are massive and lepton flavor mixing exists. Hence, I first present our current knowledge on neutrino masses and some related issues.

\subsection{Neutrino Mass Odering}
Both quarks and charged leptons take on strong mass hierarchies, namely, $m^{}_u \ll m^{}_c \ll m^{}_t$, $m^{}_d \ll m^{}_s \ll m^{}_b$ and $m^{}_e \ll m^{}_\mu \ll m^{}_\tau$. As the running masses of quarks and charged leptons are directly related to the Yukawa coupling constants in the fundamental theory, it is meaningful to make a comparison among them at a common energy scale, e.g., the Fermi scale $M^{}_Z = 91.2~{\rm GeV}$. Using the latest values of running masses of charged fermions at $M^{}_Z$ from Ref.~\cite{Xing:2011aa}, one can obtain the quark mass ratios
\begin{eqnarray}
\frac{m^{}_u}{m^{}_c} \approx 2.2^{+1.1}_{-0.8} \cdot 10^{-3} \; , ~~~ \frac{m^{}_c}{m^{}_t} \approx 3.7^{+0.3}_{-0.5} \cdot 10^{-3} \; , ~~~
\frac{m^{}_d}{m^{}_s} \approx 4.9^{+2.4}_{-1.8} \cdot 10^{-2} \; , ~~~
\frac{m^{}_s}{m^{}_b} \approx 2.0^{+0.7}_{-0.5} \cdot 10^{-2} \; , ~~~
\end{eqnarray}
and the charged-lepton ones $m^{}_e/m^{}_\mu \approx 4.7\cdot 10^{-3}$ and $m^{}_\mu/m^{}_\tau \approx 5.9 \cdot 10^{-2}$, where the uncertainties for charged-lepton mass ratios are negligible. But for neutrinos, it is not yet determined whether their mass ordering is normal (i.e., NO for $m^{}_1 < m^{}_2 < m^{}_3$) or inverted (i.e., IO for $m^{}_3 < m^{}_1 < m^{}_2$). The accelerator (T2K, NO$\nu$A and LBNF/DUNE), reactor (JUNO and RENO-50), atmospheric (PINGU, ORCA, INO and Hyper-Kamiokande) neutrino oscillation experiments are able to finally resolve this problem~\cite{Long:LP15}.

It seems plausible that neutrino mass ordering is normal, as we have seen in the case of charged fermions. Furthermore, if neutrino masses are also hierarchical, we get $m^{}_1/m^{}_2 \approx 0$ and $m^{}_2/m^{}_3  \approx (\Delta m^2_{21}/\Delta m^2_{31})^{1/2} \approx 0.17$, where $\Delta m^2_{21} \equiv m^2_2 - m^2_1 = 7.5\times 10^{-5}~{\rm eV}^2$ and $\Delta m^2_{31} \equiv m^2_3 - m^2_1 = 2.5\times 10^{-3}~{\rm eV}^2$ extracted from neutrino oscillation experiments have been used. Although the lightest neutrino mass ($m^{}_1$ for NO or $m^{}_3$ for IO) is still allowed to be zero, the neutrino mass hierarchy in either NO or IO case is not following the strongly hierarchial pattern of charged-fermion masses. It is also possible that neutrino masses are nearly degenerate, i.e., $m^{}_1/m^{}_2 \approx m^{}_2/m^{}_3 \approx 1$, which are completely different from the charged-fermion mass ratios. Thus, the determination of neutrino mass ordering and the absolute mass scale are of crucial importance to achieve a unified description of fermion mass spectra and reveal the underlying symmetry between quarks and leptons.

\subsection{Absolute Neutrino Masses}

Currently there are three practical ways to constrain absolute neutrino masses. The first one is to measure precisely the energy spectrum of electrons from tritium beta decays, particularly in the region close to the endpoint. From the observed spectrum, one can extract the information on the effective neutrino mass $m^{}_\beta \equiv \sqrt{|U^{}_{e1}|^2 m^2_1 + |U^{}_{e2}|^2 m^2_2 +|U^{}_{e3}|^2 m^2_3}$, where $U^{}_{ei}$ for $i = 1, 2, 3$ are the first-row elements of the PMNS matrix. The Mainz and Troitsk experiments have set an upper limit on the effective neutrino mass $m^{}_\beta < 2.2~{\rm eV}$  at the $95\%$ confidence level (C.L.)~\cite{Kraus:2004zw}. In the near future, the KATRIN experiment will improve this upper limit by one order of magnitude, namely, $m^{}_\beta < 0.2~{\rm eV}$~\cite{Wolf:2008hf}. 

Second, useful constraints on the absolute scale of neutrino masses can also be obtained from the experimental searches for neutrinoless double-beta decays $N(Z, A) \to N(Z+2, A) + 2e^-$, in which the effective neutrino mass $m^{}_{\beta \beta} \equiv |U^2_{e1} m^{}_1 + U^2_{e2} m^{}_2 + U^2_{e3} m^{}_3|$ represents the contributions from light-neutrino exchanges. The upper bound reported by the EXO Collaboration~\cite{Albert:2014awa} is $m^{}_{\beta \beta}  < (0.20 \cdots 0.69)~{\rm eV}$ at the $90\%$ C.L.,  while that by the KamLAND-Zen experiment~\cite{Gando:2012zm} is $m^{}_{\beta \beta} < (0.15\cdots 0.52)~{\rm eV}$, where the wide ranges are caused by  the uncertainties in the evaluation of nuclear matrix elements assosciated with the ${^{136}}{\rm Xe}$ nuclei. In addition, the $^{76}{\rm Ge}$-based experiment GERDA gives $m^{}_{\beta \beta} < (0.22 \cdots 0.64)~{\rm eV}$ at the same C.L.~\cite{Agostini:2013mzu}. For a recent review, see Ref.~\cite{Bilenky:2014uka}.

\begin{figure}[t]
\centering
\includegraphics[height=3.0in]{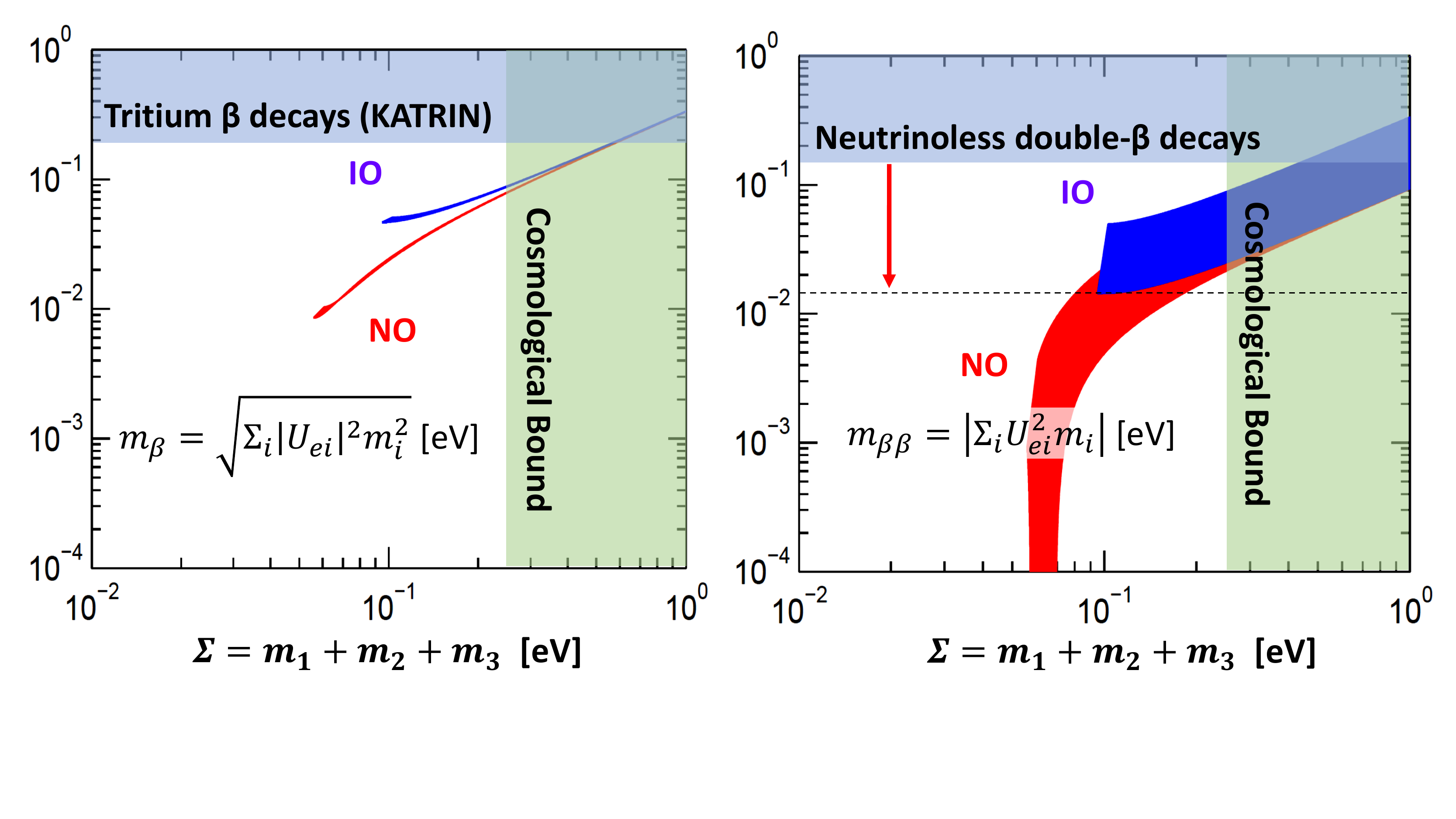}
\vspace{-1.5cm}
\caption{\it The allowed ranges of the effective neutrino masses in beta decays $m^{}_\beta$ and neutrinoless double-beta decays $m^{}_{\beta \beta}$, where current and future experimental constraints are indicated by shaded areas.}
\label{fig:absolute}
\end{figure}

Third, the precise measurements of cosmic microwave background and large-scale structure formation are sensitive to the sum of neutrino masses $\Sigma \equiv m^{}_1 + m^{}_2 + m^{}_3$. The latest observation by the Planck Collaboration, together with Baryon Accoustic Oscillations, leads to $\Sigma < 0.23~{\rm eV}$ at the $95\%$ C.L.~\cite{Ade:2015xua}. However, the cosmological bound actually depends on the chosen data sets and the model of cosmology~\cite{Lesgourgues:2012uu}.  In the $\Lambda$CDM cosmology, the future experiments will be able to reach a sensitivity of $\sigma(\Sigma) = 0.016~{\rm eV}$~\cite{Abazajian:2013oma}, which is accurate enough to observe a positive signal of nonzero neutrino masses and even sensitive to neutrino mass ordering, as shown in Fig.~\ref{fig:absolute}
\subsection{Type of Neutrino Masses}

Massive neutrinos can be either Dirac or Majorana particles. In the Dirac case, both neutrinos and antineutrinos have left-handed and right-handed components. However, the right-handed neutrinos $\nu^{}_{\rm R}$ are singlets under all the SM gauge symmetries, so one has to enforce the conservation of lepton number in order to forbid a Majorana mass term $M^{}_{\rm R} \overline{\nu^C_{\rm R}} \nu^{}_{\rm R}$. The lepton number is accidentally conserved in the SM at the classical level, but is anomalously violated. Furthermore, for Dirac neutrinos, the smallness of their masses requires extremely tiny Yukawa couplings, which should be over twelve orders of magnitude smaller than the top-quark Yukawa coupling. This exaggerates the strong hierarchy problem of fermion masses.

In the Majorana case, lepton number is no longer a good quantum number, and it is unnecessary to distinguish between neutrinos and antineutrinos. Both left-handed and right-handed light neutrinos take part in the weak interactions, and are produced  together with charged leptons and anti-leptons via the charged-current interaction, respectively. As we shall discuss later, light Majorana neutrinos can be realized in a class of seesaw models. 

At present, only the neutrinoless double-beta decays are a feasible way to demonstrate the Majorana nature of massive neutrinos. Nevertheless, if neutrino mixing parameters happen to be in a special region such that significant cancellation takes place in $m^{}_{\beta \beta}$, then the rate of neutrinoless double-beta decays will be highly suppressed. In this case, it becomes impossible to make a decisive conclusion on Dirac or Majorana nature of neutrinos in the near future. 

\subsection{Flavor Mixing and CP Violation}

One can find the latest global-fit analysis of neutrino oscillation data in Refs.~\cite{Capozzi:2013csa}. Although three different groups have found distinct best-fit values of neutrino mixing parameters, in particular $\theta^{}_{23}$ and the Dirac CP-violating phase $\delta$, their results are perfectly consistent with each other at the $3\sigma$ level. The determinations of neutrino mass ordering, the octant of $\theta^{}_{23}$, and  $\delta$ are the primary goals of ongoing and forthcoming neutrino oscillation experiments. It is worth mentioning that there exists a weak hint on a maximal CP-violating phase $\delta \sim 270^\circ$, which is very promising for a direct discovery of leptonic CP violation in the foreseeable future. 

With the help of the global-fit results, one can figure out the allowed ranges of the absolute values of PMNS matrix elements~\cite{Capozzi:2013csa}
\begin{eqnarray}
\left(\matrix{|U^{}_{e1}| & |U^{}_{e2}| & |U^{}_{e3}| \cr |U^{}_{\mu 1}| & |U^{}_{\mu 2}| & |U^{}_{\mu 3}| \cr |U^{}_{\tau 1}| & |U^{}_{\tau 2}| & |U^{}_{\tau 3}|}\right) = \left(\matrix{0.801\cdots 0.845 & 0.514\cdots 0.580 & 0.137\cdots 0.158 \cr 0.225\cdots 0.517 & 0.441\cdots 0.699 & 0.614\cdots 0.793 \cr 0.246\cdots 0.529 & 0.464\cdots 0.713 & 0.590\cdots 0.776 }\right) \; ,
\end{eqnarray}
from which we can observe a possible $\mu$-$\tau$ symmetry $|U^{}_{\mu i}| = |U^{}_{\tau i}|$ for $i = 1, 2, 3$~\cite{Fukuyama:1997ky}. If the $\mu$-$\tau$ symmetry is preserved, the flavor mixing angles are restrictively constrained~\cite{Xing:2008fg}: (1) $\theta^{}_{23} = 45^\circ$ and $\theta^{}_{13} = 0$; (2) $\theta^{}_{23} = 45^\circ$ and $\delta = 90^\circ$ or $270^\circ$. Since $\theta^{}_{13} = 0$ has already been excluded by the reactor neutrino experiments, we are left with only the second possibility. It is interesting that the weak hint on $\delta \sim 270^\circ$ coincides with the $\mu$-$\tau$ symmetry. Although a maximal mixing angle $\theta^{}_{23} = 45^\circ$ is compatible with oscillation experiments within $1\sigma$, the best fit to oscillation data points to $\theta^{}_{23} \neq 45^\circ$. If such a deviation is indeed confirmed, we expect that a partial $\mu$-$\tau$ symmetry with $|U^{}_{\mu 1}| = |U^{}_{\tau 1}|$ or $|U^{}_{\mu 2}| = |U^{}_{\tau 2}|$ exists in the PMNS matrix~\cite{Xing:2014zka}.

Apart from neutrino mass ordering, leptonic CP violation in neutrino oscillations is the main task for future oscillation experiments~\cite{Branco:2011zb}. The Dirac CP phase $\delta$ can be extracted from the difference between neutrino and antineutrino oscillation probabilities, although the MSW matter effects induce fake CP violation in the long-baseline accelerator and atmospheric neutrino experiments. For Majorana neutrinos, one ultimately unavoidable question is how to probe two Majorana CP-violating phases. Neutrinoless double-beta decays, neutrino-antineutrino oscillations~\cite{Pontecorvo:1957cp,Schechter:1980gk} and other related lepton-number-violating processes could provide a clue. But robust evidence for Majorana nature of massive neutrinos should first be found in the neutrinoless double-beta decay experiments, rendering the determination of Majorana CP phases necessary.

\subsection{Electromagnetic Properties of Neutrinos}

Another issue associated with massive neutrinos is the electromagnetic properties~\cite{Giunti:2014ixa}. If the SM is extended to accommodate massive Dirac neutrinos, the magnetic dipole moment of neutrinos can be calculated $\mu^{}_{\nu_i} = 3\times 10^{-20} (m^{}_i/0.1~{\rm eV}) \mu^{}_{\rm B}$, where $\mu^{}_{\rm B} \equiv e/2m^{}_e$ is the Bohr magneton~\cite{Fujikawa:1980yx}. For Majorana neutrinos, the magnetic moments turn out to be vanishing due to the neutrality condition $\nu^{}_i = \nu^C_i$. However, the transitional electric $\epsilon^{}_{ij}$ and magnetic $\mu^{}_{ij}$ dipole moments can be nonzero for both Dirac and Majorana neutrinos. Because of transitional electromagnetic dipole moments, neutrinos become unstable and decay via $\nu^{}_i \to \nu^{}_j + \gamma$ for $m^{}_i > m^{}_j$. Moreover, they can interact with external electromagnetic fields, and electrons with additional contributions from the electromagnetic vertex.

The implications of neutrino electromagnetic interactions for particle physics and astrophysics lead to stringent bounds on the effective dipole moment $\mu^{}_{\rm eff} = \sqrt{|\mu^{}_{ij}|^2 + |\epsilon^{}_{ij}|^2}$. First, the elastic (anti)neutrino-electron scattering receives the electromagnetic contribution in addition to the SM one. The best upper limit from this kind of laboratory experiments is $\mu^{}_{\rm eff} < 2.9\times 10^{-11} \mu^{}_{\rm B}$ from the GEMMA Collaboration~\cite{Beda:2012zz}. Second, the plasmon decay into neutrino-antineutrino pairs dominates the neutrino production in white dwarfs or the cores of globular-cluster red giants. The requirement of no extra energy losses via neutrinos in the globular cluster gives rise to the most restictive limit $\mu^{}_{\rm eff} \lesssim 3.0\times 10^{-12} \mu^{}_{\rm B}$~\cite{Raffelt:1989xu}.

\section{Origin of Neutrino Masses and Flavor Mixing}
\begin{figure}[t]
\centering
\includegraphics[height=3.3in]{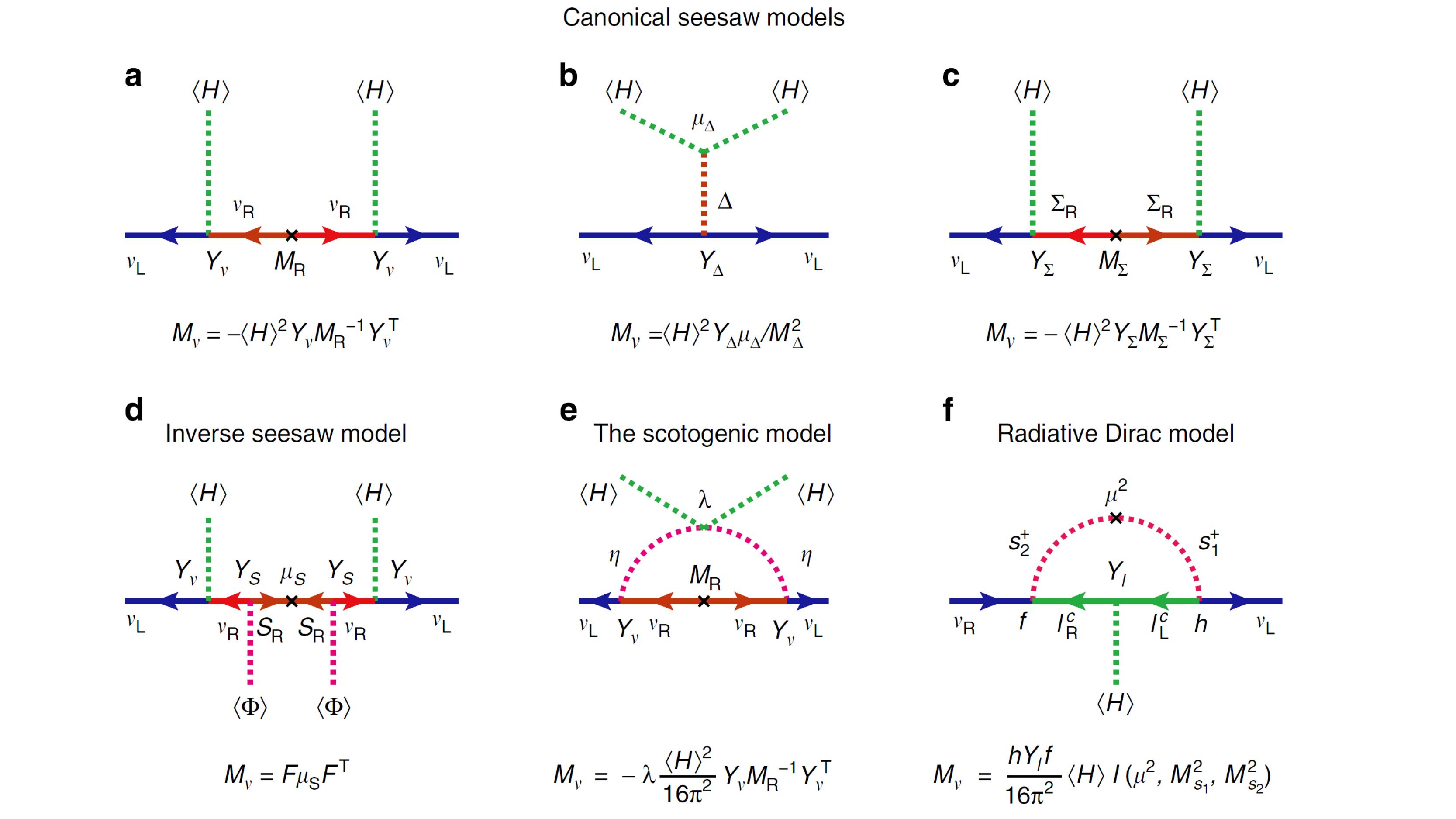}
\caption{\it An incomplete list of neutrino mass models, where the generated neutrino mass matrix has been given below the corresponding Feynman diagram~\cite{Ohlsson:2013xva}.}
\label{fig:seesaw}
\end{figure}
Now that neutrinos are massive particles, one immediate question is how to extend the SM and generate tiny neutrino masses in a natural way. Another important question is why flaovr mixing angles in the lepton sector are so different from those in the quark sector. In the following, I introduce neutrino mass models and the approaches implemented to explain lepton flavor mixing.

\subsection{Canonical Seesaw Models}

As pointed out by Weinberg, one can regard the SM as an effective theory at the electroweak scale, and introduce a dimension-five operator $(\overline{\ell^{}_{\rm L}} \cdot \tilde{H}) (\tilde{H}^T \cdot \ell^C_{\rm L})$ to generate tiny neutrino masses of Majorana type. In this case, the smallness of neutrino masses can be ascribed to the existence of a new high energy or mass scale. See, e.g., Fig.~\ref{fig:seesaw}, for a partial list of neutrino mass models. Since both $\ell^{}_{\rm L}$ and $\tilde{H}$ belong to the $({\bf 2}, -{\bf 1})$ representation of the $SU(2)^{}_{\rm L} \times U(1)^{}_{\rm Y}$ gauge symmetry, there are three distinct ways to construct a renormalizable full theory at a superhigh-energy scale:
\begin{itemize}
\item Type-I Seesaw~\cite{Minkowski:1977sc}: Three right-handed neutrino singlets $\nu^{}_{\rm R} \sim ({\bf 1}, {\bf 0})$ are coupled to both lepton and Higgs doublets $ Y^{}_\nu \overline{\ell^{}_{\rm L}}  H \nu^{}_{\rm R}$, where $Y^{}_\nu$ is the Yukawa coupling matrix. Now that $\nu^{}_{\rm R}$ are gauge singlets, a Majorana mass term $M^{}_{\rm R} \overline{\nu^C_{\rm R}} \nu^{}_{\rm R}$ is allowed. The Majorana mass matrix of light neutrinos is given by $M^{}_\nu = - \langle H \rangle^2 Y^{}_\nu M^{-1}_{\rm R} Y^T_\nu $ for ${\cal O}(M^{}_{\rm R}) \gg \langle H \rangle \approx 174~{\rm GeV}$.

\item Type-II Seesaw~\cite{type2}: The SM is extended with a Higgs triplet $\Delta \sim ({\bf 3}, -{\bf 2})$, which is coupled to lepton doublets via $\frac{1}{2} Y^{}_\Delta \overline{\ell^{}_{\rm L}} \Delta i\sigma^{}_2 \ell^C_{\rm L}$  and to Higgs doublets $\mu^{}_\Delta H^T i\sigma^{}_2 \Delta H$. Light neutrinos acquire a tiny Majorana mass term with $M^{}_\nu = Y^{}_\Delta v^{}_\Delta$ after the spontaneous symmetry breaking $\langle \Delta \rangle = v^{}_\Delta \approx \mu^{}_\Delta \langle H \rangle^2/M^2_\Delta \ll \langle H \rangle$, where $M^{}_\Delta$ is the Higgs triplet mass.

\item Type-III Seesaw~\cite{type3}: Three triplet fermions $\Sigma^{}_{\rm R} \sim ({\bf 3}, {\bf 0})$ are coupled to both lepton and Higgs doublets via $Y^{}_\Sigma \overline{\ell^{}_{\rm L}} \Sigma^{}_{\rm R} \tilde{H}$, and light neutrino masses are given by $M^{}_\nu = - \langle H \rangle^2 Y^{}_\Sigma M^{-1}_\Sigma Y^T_\Sigma$, where $M^{}_\Sigma$ is the fermion triplet mass. In this case, the neutral component of $\Sigma^{}_{\rm R}$ plays the same role as $\nu^{}_{\rm R}$ in the type-I seesaw model.
\end{itemize}
It seems natural that the mass scale of new particles in the canonical seesaw models is close to the scale of grand unified theories (GUT's), i.e., $\Lambda^{}_{\rm GUT} \sim 10^{16}~{\rm GeV}$. At the low-energy scale, one can integrate heavy particles out and obtain the Weinberg operator.  Another salient feature of seesaw models is to explain the matter-antimatter asymmetry via leptogenesis~\cite{Fukugita:1986hr}.

However, such a high-scale seesaw model suffers from two potential problems: (1) The new particles introduced in canonical seesaw models are so heavy that they cannot be directly tested in terrestrial experiments; (2) The radiative corrections to the Higgs mass turn out to be large, causing the hierarchy or naturalness problem~\cite{Vissani:1997ys}. Possible solutions to those problems are either to seek a seesaw scenario in the TeV region, or to generate neutrino masses in a completely different way. In fact, the ATLAS and CMS collaborations at the CERN Large Hadron Collider have performed dedicated searches for lepton-number-violating signals in the low-scale Type-I~\cite{Khachatryan:2015gha}, Type-II~\cite{Chatrchyan:2012ya} and Type-III~\cite{Aad:2015dha} seesaw models, and set restrictive limits on the model parameters. 

\subsection{Radiative Mass Models}

In canonical seesaw models, the suppression of neutrino masses is achieved by the introduction of a high energy or mass scale. In contrast, tiny neutrino masses can arise from loop corrections, as in a class of radiative mass models~\cite{Zee:1980ai}. In the latter case, new heavy particles running in the loops can be around or even below the TeV scale, when the relevant coupling constants are reasonably small rather than finely tuned. Two typical examples are given in the last two diagrams in Fig.~\ref{fig:seesaw}. Now neutrino masses are actually suppressed by a loop factor and small dimensionless couplings. An intriguing feature of radiative mass models is the possible connection to neutrino masses, flavor symmetries, collider signals and dark matter, implying a very rich phenomenology.

Instead of presenting a concrete example, I make some comments on radiative neutrino mass models in the scale-invariant extension of the SM~\cite{Lindner:2014oea}. The main motivation to consider a scale-invariant extension is to solve the gauge hierarchy problem of the SM. The Higgs mass or the electroweak scale gets huge corrections from new physics at a superhigh-energy scale, for instance, the Planck scale. As demonstrated by Coleman and Weinberg in a $U(1)$ gauge theory of scalar fields~\cite{Coleman:1973jx}, if all the parameters of mass dimension are eliminated, the theory becomes classically scale-invariant and the  spontaneous gauge symmetry breaking can be triggered by radiative corrections. Consequently, a mass scale emerges and only a logarithmic scale dependence of mass parameters is left. Along this line, one can construct a radiative neutrino mass model by using only dimensionless couplings and the vacuum expectation values of scalar fields come in as mass scales in the theory~\cite{Lindner:2014oea}.

\subsection{Discrete Flavor Symmetries}

The mechanisms for neutrino mass generation are usually not responsible for the lepton flavor mixing pattern. In the past decade, a great number of non-Abelian discrete flavor symmetries, such as $S^{}_3$, $A^{}_4$, $S^{}_4$, $A^{}_5$, $T^\prime$ and $\Delta(27)$, have been implemented to explain the observed lepton flavor mixing~\cite{Altarelli:2010gt}. A paradigm for model building is based on $A^{}_4$, the symmetry group of a tetrahedron~\cite{Ma:2001dn}. This was particularly interesting when neutrino oscillation data were favoring the so-called tribimaximal (TB) mixing~\cite{Harrison:2002er}
\begin{equation}
U^{}_{\rm TB} = \left(\matrix{\displaystyle \frac{2}{\sqrt{6}} & \displaystyle \frac{1}{\sqrt{3}} & 0 \cr \displaystyle -\frac{1}{\sqrt{6}} & \displaystyle \frac{1}{\sqrt{3}} & \displaystyle \frac{1}{\sqrt{2}} \cr \displaystyle \frac{1}{\sqrt{6}} & \displaystyle -\frac{1}{\sqrt{3}} & \displaystyle \frac{1}{\sqrt{2}}}\right) \; ,
\end{equation}
resulting in $\theta^{}_{12} = 35.3^\circ$, $\theta^{}_{23} = 45^\circ$, $\theta^{}_{13} = 0$ and trivial CP-violating phases. The essential idea is to impose a global flavor symmetry $G^{}_f$ on the generic Lagrangian, and the symmetry is broken into residual symmetries $G^{}_l$ and $G^{}_\nu$ in the charged-lepton and neutrino sectors, respectively. It is the difference between the symmetry breaking patterns of $G^{}_f \to G^{}_l$ and $G^{}_f \to G^{}_\nu$ that leads to a nontrivial PMNS matrix.

However, the prediction of $\theta^{}_{13} = 0$ from the TB mixing pattern has already been excluded by the Daya Bay reactor neutrino experiment~\cite{An:2012eh}. Hence the flavor models based on $A^{}_4$ and other discrete symmetries need to be considerably changed to accommodate a relatively large $\theta^{}_{13}$. More importantly, a nonzero $\theta^{}_{13}$ is a prerequisite for leptonic CP violation, so it is intriguing to construct a flavor model to account for both $\theta^{}_{13}$ and the Dirac CP-violating phase $\delta$. Regarding this point, the $\mu$-$\tau$ reflection symmetry has to be mentioned~\cite{Harrison:2002et}. In the basis where the charged-lepton mass matrix $M^{}_\ell$ is diagonal, the neutrino mass matrix $M^{}_\nu$ that is invariant under $\nu^{}_e \to \nu^C_e$, $\nu^{}_\mu \to \nu^C_\tau$ and $\nu^{}_\tau \to \nu^C_\mu$ leads to $\theta^{}_{23} = 45^\circ$ and $\delta = 90^\circ$ or $270^\circ$, while $\theta^{}_{12}$ and $\theta^{}_{13}$ are arbitrary. The $\mu$-$\tau$ reflection symmetry is actually a special case of the generalized CP symmetry, which combines the ordinary CP transformation and a discrete flavor symmetry~\cite{Mohapatra:2012tb}. To be explicit, given the fields $\varphi(x)$, the generalized CP transformation is defined by $\varphi(x) \to X^{}_{\bf r} \varphi^*(x^\prime)$, where $x^\prime = (t, -{\bf x})$ and $X^{}_{\bf r}$ is the matrx of transformations associated with the fields in the irreducible representation ${\bf r}$ of the discrete flavor symmetry $G^{}_f$. If the consistency condition $\rho^{}_{\bf r}(g) = X^{}_{\bf r} \rho^*_{\bf r}(g^\prime) X^{-1}_{\bf r}$ is satisfied, 
the generalized CP symmetry and discrete flavor symmetry can be integrated into a full symmetry. The predicitions for CP violation are then dependent on how the full symmetry is broken~\cite{Feruglio:2013hia}.

Finally, when the flavor symmetry is brought into the GUT framework, a possible connection of $\theta^{}_{13}$ to the Cabibbo angle $\theta^{}_{\rm C}$ can be established~\cite{Antusch:2011qg}. In the quark sector, strong mass hierarchies indicate that the dominant mixing angle stems from the down-type quark mass matrix $M^{}_d$, namely, $\theta^{\rm d}_{12} \approx \sqrt{m^{}_d/m^{}_s} \approx \theta^{}_{\rm C}$. On the other hand, the down-type quark mass matrix is related to the charged-lepton one $M^{}_\ell$ in GUT's, so we have $\theta^{\ell}_{12} \approx \alpha^{}_{12} \theta^{\rm d}_{12} \approx \alpha^{}_{12} \theta^{}_{\rm C}$ where $\alpha^{}_{12}$ is fixed by the Clebsch factors. If the flavor symmetry is further utilized to gurantee the TB mixing for neutrinos $U^{}_\nu = U^{}_{\rm TB}$, $\theta^{}_{13}$ in the PMNS matrix receives the contributions from both charged-lepton and neutrino sectors, i.e., $\theta^{}_{13} \approx \theta^{}_{\rm C}/\sqrt{2}$. In addition, a correlation among neutrino mixing angles and CP-violating phase $\theta^{}_{12} = \theta^\nu_{12} + \theta^{}_{13} \cos \delta$ can also be derived~\cite{Antusch:2011qg}, where $\theta^\nu_{12} = 35.3^\circ$ as predicted by the TB mixing.

\subsection{Texture Zeros}

In most models of non-Abelian discrete flavor symmetries, the flavor mixing angles are in general decoupled from the lepton masses. In contrast, the texture zeros in fermion mass matrices usually imply the relationship among flavor mixing angles and fermion mass ratios. In particular, the texture zeros in quark mass matrices give rise to $\theta^{}_{\rm C} \approx \sqrt{m^{}_d/m^{}_s}$, which is in perfect agreement with experimental observations~\cite{Weinberg:1977hb}. However, as we have mentioned before, the mass hierarchy of neutrinos is not as strong as that of charged fermions. The weak neutrino mass hierarchy may be associated with large neutrino mixing angles.

 In the flavor basis where the charged-lepton flavor eigenstates coincide with their mass eigenstates, the Majorana neutrino mass matrix $M^{}_\nu$ possesses six independent complex elements, which can be reconstructed from three mixing angles, three mass eigenvalues and three CP-violating phases. If two independent matrix elements are taken to be zero, there are four real constraints among neutrino mixing parameters~\cite{Frampton:2002yf}. As a consequence, one can determine the absolute neutrino mass scale and three CP phases in terms of precisely measured three neutrino mixing angles and two neutrino mass-squared differences. Currently, seven two-zero textures survive neutrino oscillation data, and will be tested by the precision measurements of neutrino mixing parameters. 

It should be noticed that the approaches of texture zeros are intimately related to those of discrete flavor symmetries. Both Abelian and non-Abelian symmetries can be used to realize texture zeros in the lepton mass matrices in the cases either with or without seesaw mechanisms.

\section{Recent Progress on Astrophysical Neutrinos}

In the last part of my talk, I turn to the recent progress in the studies of astrophysical neutrinos, including the keV sterile neutrinos, supernova neutrinos and ultrahigh-energy cosmic neutrinos. A particular emphasis will be placed on how to investigate the fundamental properties of neutrinos in astrophysical environments.  

\subsection{keV Sterile Neutrinos}

In 2014, two independent groups discovered an X-ray line around 3.55 keV in the dark-matter dominated astrophysical objects~\cite{Bulbul:2014sua}. The first evidence comes from the stacking X-ray spectra of central parts of 81 galaxy clusters, which have been observed by XMM-Newton and Chandra telescopes. The second one is found in the nearby Andromeda galaxy, the Perseus cluster and the new blank-sky dataset observed by XMM-Newton. For each group with different datasets, the global significance of the X-ray line observation is above $4\sigma$. One possible interpretation is the radiative decays of a sterile neutrino of 7.1 keV mass $\nu^{}_s \to \nu^{}_a + \gamma$, where $\nu^{}_a$ denotes an active neutrino. These decays take place if there exists a sterile-active neutrino mixing and active neutrinos decay radiatively at an extremely small rate. The sterile-active mixing angle $\theta^{}_{\rm s}$ can be determined from the observed strength of the X-ray line, implying $\sin^2 2\theta^{}_{\rm s} = 4.9\times 10^{-11}$~\cite{Bulbul:2014sua}. 

If this X-ray line is confirmed by future observations, one has to embed the keV sterile neutrino into the theory of massive neutrinos. In the type-I seesaw model, the lightest right-handed neutrino can be a good candidate for the 7.1 keV-mass sterile neutrino~\cite{Boyarsky:2009ix}. In this case, active neutrino masses receive the contributions only from two much heavier right-handed neutrinos, as the mixing angle between keV sterile neutrinos and active neutrinos is extremely small. On the other hand, the production of keV sterile neutrinos should be aided by resonant MSW effects, which require a remarkable primordial lepton asymmetry~\cite{Shi:1998km}, or by some other mechanisms.  

\subsection{Supernova Neutrinos}

The flavor conversions of supernova neutrinos are complicated by the interaction of neutrinos with dense background particles, which has attracted a lot of attention~\cite{Duan:2009cd}. In the supernova envelope, where the matter density is appropriate for neutrinos to experience the MSW resonances corresponding to two neutrino mass-squared differences. Further inward, both neutrino and matter densities become very large. While the dense matter tends to suppress the mixing angle, the dense neutrino gas induces nonlinear refraction effects via the self-interaction of neutrinos~\cite{Pantaleone:1992eq}. 

In a dense neutrino gas, neutrinos of different energies are coupled together via neutrino-neutrino interaction and they could oscillate cooperatively over a wide range of energies. Collective oscillations of supernova neutrinos have been found to considerably modify neutrino energy spectra~\cite{Duan:2009cd}. For instance, in the two-flavor approximation, $\nu^{}_e$ and $\nu^{}_\tau$ can exchange their spectra above a critical energy, below which neutrino flavor conversions never happen. Depending on the initial neutrino spectra, multiple spectral splits are also possible. The conditions for the flavor instability have been explored in assumption of spherical symmetry about the supernova center and axial symmetry about the radial direction. 

The latest development in this area is the discovery of spontaneous symmetry breaking, which means that the inital conditions respect a presumed symmetry but the equations of motion allow for symmetry-breaking solutions~\cite{Raffelt:2013rqa}. For now, it remains to see whether the flavor instability is reached in a real supernova environment.

\subsection{Ultrahigh-energy Cosmic Neutrinos}

The ${\rm km}^3$-scale neutrino telescope, IceCube, has recently reported the detection of 37 neutrino events in the energy range from 28 TeV to 2 PeV, among which 28 events are identified as electromagnetic or hadronic cascades while 9 events as muon tracks. The hypothesis of only atmospheric neutrino background is already exlcuded at the $5.7\sigma$ level. Assuming a diffusive astrophysical source, one can obtain a power-law index of $\alpha = 2.3 \pm 0.3$ for the energy spectrum $E^{-\alpha}$ and a flux of $10^{-8}~{\rm GeV}~{\rm cm}^{-2}~{\rm s}^{-1}~{\rm sr}^{-1}$ per flavor. The flavor composition at the detector is consistent with $\phi^{}_e : \phi^{}_\mu : \phi^{}_\tau = 1 : 1 : 1$, as predicted by an initial flavor ratio $\phi^0_e : \phi^0_\mu : \phi^0_\tau = 1 : 2 : 0$ and modified by neutrino oscillations~\cite{Aartsen:2014gkd}.

The origin of ultrahigh-energy cosmic neutrinos could be astrophysical objects, such as Active Galactic Nuclei, Gamma-Ray Bursts and Starbust Galaxies~\cite{Anchordoqui:2013dnh}. Here I focus on how to constrain the secret neutrino interactions with the help of these high-energy neutrinos~\cite{Ng:2014pca}. Suppose that the neutrino-neutrino interaction with a coupling constant $g$ is mediated by a light scalar particle $\phi$, whose mass $m^{}_\phi$ varies from keV to GeV. Requiring that the ultrahigh-energy cosmic neutrinos are not remarkably absorbed by the cosmic neutrino background, one can derive stringent limits on $g$ over $m^{}_\phi \in [1, 10]$ MeV. These limits complement those from Supernova 1987A, Big Bang Nucleosynthesis, and Cosmic Microwave Background~\cite{Kolb:1987qy}.

\section{Summary and Outlook}

Over half a century, our knowledge about the fundamental properties of neutrinos has been greatly extended. The most important information is that neutrinos are massive particles and lepton flavors are mixed, as firmly established in neutrino oscillation experiments.  In the foreseeable future, neutrino mass ordering, leptonic CP violation, the Majorana nature of neutrinos and the absolute neutrino mass scale will hopefully be fixed. With this tremendous progress and further information provided by astrophysics and cosmology, we expect that a complete theory of massive neutrinos and lepton flavor mixing will emerge eventually.

\section*{Acknowledgements}

I would like to thank Prof. Peter Krizan and Prof. Marko Mikuz for their kind invitation to give a review talk at this conference, and Prof. Zhi-zhong Xing, Dr. Jue Zhang, Dr. Zhen-Hua Zhao and Dr. Ye-Ling Zhou for their great help in preparing this talk.

\end{document}